\documentclass[superscriptaddress,amssymb,amsmath,nobibnotes,aps,prd,showpacs,
nofootinbib,twocolumn]{revtex4}%
\usepackage[utf8x]{inputenc}
\usepackage{graphicx}
\usepackage{epsf}
\usepackage{bm}
\usepackage{amsmath}
\usepackage{amsfonts}
\usepackage{amssymb}
\usepackage{epstopdf}%
\providecommand{\U}[1]{\protect\rule{.1in}{.1in}}

\newcommand{\be}{\begin{equation}}
\newcommand{\ee}{\end{equation}}
\newcommand{\bear}{\begin{eqnarray}}
\newcommand{\eear}{\end{eqnarray}}

\newcommand{\mincir}{\raise
-3.truept\hbox{\rlap{\hbox{$\sim$}}\raise4.truept\hbox{$<$}\ }}
\newcommand{\magcir}{\raise
-3.truept\hbox{\rlap{\hbox{$\sim$}}\raise4.truept\hbox{$>$}\ }}

\usepackage{cancel}
\usepackage[colorlinks]{hyperref}
\usepackage[usenames,dvipsnames]{color}
\hypersetup{
     breaklinks=true,
    pdfstartview={FitH},    
    colorlinks=true,       
    linkcolor=blue,          
    citecolor=red,        
    filecolor=magenta,      
    urlcolor=blue,           
    anchorcolor=green,      
    linktocpage=true
}

\begin{document}

\title{Dynamical system analysis of Myrzakulov gravity}
\author{G. Papagiannopoulos }
\affiliation{Department of Physics, National \& Kapodistrian University of Athens, Zografou
Campus GR 157 73, Athens, Greece}
\author{Spyros Basilakos}
\affiliation{National Observatory of Athens, Lofos Nymfon, 11852 Athens, Greece}
\affiliation{Academy of Athens, Research Center for Astronomy and Applied Mathematics,
Soranou Efesiou 4, 11527, Athens, Greece}
\affiliation{
School of Sciences, European University Cyprus, Diogenes Street, Engomi 1516 
Nicosia}
\author{Emmanuel N. Saridakis}
\affiliation{National Observatory of Athens, Lofos Nymfon, 11852 Athens, Greece}
\affiliation{CAS Key Laboratory for Researches in Galaxies and Cosmology, Department of
Astronomy, University of Science and Technology of China, Hefei, Anhui 230026,
P.R. China}
\affiliation{Departamento de Matem\'{a}ticas, Universidad Cat\'{o}lica del 
Norte, 
Avda.
Angamos 0610, Casilla 1280 Antofagasta, Chile}

\begin{abstract}
We perform a dynamical   system analysis of Myrzakulov  or $F(R,T)$  
 gravity, which is a subclass of affinely connected metric theories, where  
ones uses a specific but non-special connection that allows for 
non-zero curvature and torsion simultaneously. We consider two classes of 
models, we extract the critical points, and we examine their stability 
properties alongside their physical features. In  the Class 1 models, which 
possess $\Lambda$CDM cosmology as a limit, we find  the sequence 
of matter and dark energy eras, and we show that  the Universe will result in a 
dark-energy dominated critical point for  which dark energy behaves like a 
cosmological constant. Concerning the   dark-energy 
equation-of-state parameter   we   find that  it lies in the quintessence or
phantom regime, according to the value of the model parameter. 
For the Class 2 models, we again find the dark-energy dominated, de Sitter 
late-time attractor, although the scenario  does not possess $\Lambda$CDM 
cosmology as a limit. The cosmological behavior is richer, and    the 
dark-energy sector can be quintessence-like, phantom-like,  or experience 
the phantom-divide crossing during the evolution.

\end{abstract}

\maketitle


\section{Introduction}

An increasing amount of observational data has now led to the establishment of
the standard model of cosmology, according to which the universe has passed
through two phases of accelerating expansion one at early and one at late
times. Although the latter can be explained by a cosmological constant, the
possibility of a dynamical nature, some possible tensions, as well as the
necessity for an additional description of the former phase, may ask for a
kind of modification. In general one has two ways to accomplish this. The
first is to maintain general relativity as the underlying gravitational theory
but consider extra field contents such as the dark energy sector
\cite{Copeland:2006wr,Cai:2009zp} and/or the inflaton field
\cite{Bartolo:2004if}. The second way is to construct modified gravity
theories, which in a particular limit tend to general relativity, but which in
general exhibit extra degrees of freedom that can drive the non-standard
universe evolution \cite{CANTATA:2021ktz,Capozziello:2011et,Nojiri:2010wj}. 
This direction has the additional advantage of bringing gravity closer to a 
quantum description \cite{Addazi:2021xuf}.

Modified gravity can arise from suitable extensions of the Einstein-Hilbert
action, such as in $F(R)$ gravity \cite{DeFelice:2010aj}, {\bf{in 
theories with non-minimal coupling between matter and curvature  
\cite{Bertolami:2007gv,Bertolami:2008ab}}}, in $F(G)$ gravity
\cite{Nojiri:2005jg,DeFelice:2008wz}, in Lovelock gravity
\cite{Lovelock:1971yv,Deruelle:1989fj}, in Horndeski \cite{Horndeski:1974wa} 
and generalized galileon \cite{DeFelice:2010nf,Deffayet:2011gz} gravities, etc. 
A different way to construct
gravitational modifications is to use as a base the equivalent, teleparallel
formulation of gravity \cite{Pereira,Maluf:2013gaa}, and build torsional
theories such as $F(T)$ gravity
\cite{Cai:2015emx,Ferraro:2006jd,Linder:2010py},    
theories with non-minimal coupling between matter and torsion 
\cite{Harko:2014sja,Harko:2014aja},    $F(T,T_{G})$ gravity
\cite{Kofinas:2014owa}, $F(T,B)$ gravity \cite{Bahamonde:2015zma}, teleparallel 
Horndeski  \cite{Bahamonde:2019shr}, etc. One can proceed to other 
geometrical
modifications, thus obtaining novel extended gravity theories, such as using
non-metricity \cite{Harko:2018gxr,Anagnostopoulos:2021ydo}, or constructing more 
complex structure
such as in Finsler geometry 
\cite{Bogoslovsky:1999pp,Mavromatos:2010nk,Mavromatos:2010jt}.

An alternative way to construct gravitational modifications is to alter the
connection structure of the theory, {\bf{namely the extra degrees of freedom 
will arise form the different connection instead of the different action}}. This 
was known in the framework of
metric-affine theories
\cite{Hehl:1994ue,BeltranJimenez:2012sz,Tamanini:2012mi,Iosifidis:2021pta}, as 
well as in
Finsler-like theories where the non-linear connection may bring about extra
degrees of freedom
\cite{Kouretsis:2012ys,Basilakos:2013hua,Triantafyllopoulos:2018bli,
Ikeda:2019ckp,Konitopoulos:2021eav}. In Myrzakulov or $F(R,T)$ gravity 
\cite{Myrzakulov:2012qp} (this should not be confused with the  $F(R,T)$ 
gravity where $T$ is the 
trace of the energy-momentum tensor \cite{Harko:2011kv})
ones uses a specific but non-special connection, which allows for non-zero
curvature and torsion simultaneously, which then leads to the appearance of
extra degrees of freedom that can make the theory phenomenologically viable
\cite{Saridakis:2019qwt}. As one can show, it can be expressed as a
deformation of both general relativity and its teleparallel equivalent. Hence,
this theory lies within the class of Riemann-Cartan family of theories, which
in turn belong to the general family of affinely connected metric theories
\cite{Conroy:2017yln}. Nevertheless, the theory at hand maintains zero 
non-metricity.

The cosmological applications of Myrzakulov gravity were investigated in
\cite{Myrzakulov:2012qp,Saridakis:2019qwt,Sharif:2012gz,Momeni:2011am,
Capozziello:2014bna, 
Feola:2019zqg,Iosifidis:2021kqo,Myrzakulov:2021vel,
Harko:2021tav}, while the confrontation with observational data has been 
performed in \cite{Anagnostopoulos:2020lec}. In this work we are interested in
investigating the cosmological behavior by applying the powerful method of
dynamical system analysis. Such an approach allows to extract global
information on the cosmological evolution, independently of the specific
initial conditions or the intermediate-time behavior \cite{REZA,Coley:2003mj}.
In particular, by examining the stable critical points of the autonomously
transformed cosmological equations, one can classify the infinite number of
possible evolutions into few different classes obtained asymptotically.

The plan of the work is the following. In Section \ref{Cosmologysec} we review
Myrzakulov gravity and we present the relevant cosmological equations. In
Section \ref{analysis} we perform a detailed dynamical analysis of various
scenarios in this theory, focusing on the stable late-time solutions, and we
discuss on the physical behavior. Finally, Section \ref{Conclusions} is
devoted to the Conclusions.

\section{Cosmology in Myrzakulov Gravity}

\label{Cosmologysec}

In this section  we provide the cosmological equations in a universe
governed by Myrzakulov gravity 
\cite{Myrzakulov:2012qp,Saridakis:2019qwt}. The basic feature of the theory is
the modification of the connection, {\bf{maintaining  however zero 
non-metricity}}. As it is known, choosing a general
connection $\omega^{a}_{\,\,\,bc}$ one can construct the curvature and the
torsion tensors through the expressions   \cite{Kofinas:2014owa}
\begin{align}
&  \!\!\!\!\!\!\!\!\! R^{a}_{\,\,\, b\mu\nu}\!=\omega^{a}_{\,\,\,b\nu,\mu}-
\omega^{a}_{\,\,\,b\mu,\nu} +\omega^{a}_{\,\,\,c\mu}\omega^{c}_{\,\,\,b\nu
}-\omega^{a}_{\,\,\,c\nu} \omega^{c}_{\,\,\,b\mu}\,, \label{curvaturebastard}%
\end{align}
\begin{equation}
T^{a}_{\,\,\,\mu\nu}= e^{a}_{\,\,\,\nu,\mu}-e^{a}_{\,\,\,\mu,\nu}+\omega
^{a}_{\,\,\,b\mu}e^{b}_{\,\,\, \nu} -\omega^{a}_{\,\,\,b\nu}e^{b}_{\,\,\,\mu
}\,, \label{torsionbastard}%
\end{equation}
with $e^{\,\,\, \mu}_{a}\partial_{\mu}$ the tetrad field satisfying $g_{\mu
\nu} =\eta_{ab}\, e^{a}_{\,\,\,\mu} \, e^{b}_{\,\,\,\nu}, $ with $g_{\mu\nu}$
the metric, $\eta_{ab}=\text{diag}(-1,1,...1)$, and where Greek and Latin
indices run respectively over coordinate and tangent space, and with comma
denoting differentiation.

Amongst the infinite connections, the Levi-Civita $\Gamma_{abc}$ is the only
one that by construction leads to vanishing torsion. For clarity we will use
the superscript ``LC'' to denote the curvature tensor calculated using
$\Gamma_{abc}$, i.e. $R^{(LC)a}_{\,\,\,\ \ \ \ \ \, b\mu\nu}=\Gamma
^{a}_{\,\,\,b\nu,\mu}- \Gamma^{a}_{\,\,\,b\mu,\nu} +\Gamma^{a}_{\,\,\,c\mu
}\Gamma^{c}_{\,\,\,b\nu}-\Gamma^{a}_{\,\,\,c\nu} \Gamma^{c}_{\,\,\,b\mu}$.
Similarly, imposition of the Weitzenb{\"{o}}ck connection $W_{\,\,\,\mu\nu
}^{\lambda}=e_{a}^{\,\,\,\lambda}e^{a}_{\,\,\,\mu, \nu} $ leads to zero
curvature, and the corresponding torsion tensor becomes $T^{(W)\lambda
}_{\,\,\,\ \ \ \ \ \mu\nu}=W^{\lambda}_{\,\,\,\nu\mu}- W^{\lambda}%
_{\,\,\,\mu\nu}$, where we use the superscript ``W'' to denote quantities
calculated using $W_{\,\,\,\mu\nu}^{\lambda}$. From contractions of the above
tensors one can find the Ricci scalar corresponding to the Levi-Civita
connection:
\begin{align}
&  \!\!\!\!\!\!\!\!\!\!\!\!\!\!\!\!\!\!\!\!\!\!\!\!\!\!\!\!\!\! R^{(LC)}%
=\eta^{ab} e^{\,\,\, \mu}_{a} e^{\,\,\, \nu}_{b} \left[  \Gamma^{\lambda
}_{\,\,\,\mu\nu,\lambda} - \Gamma^{\lambda}_{\,\,\,\mu\lambda,\nu}\right.
\nonumber\\
&  \ \ \ \ \ \ \ \ \ \ \left.  + \Gamma^{\rho}_{\,\,\,\mu\nu}\Gamma^{\lambda
}_{\,\,\,\lambda\rho} -\Gamma^{\rho}_{\,\,\,\mu\lambda}\Gamma^{\lambda
}_{\,\,\,\nu\rho} \right]  ,
\end{align}
as well as the torsion scalar corresponding to the Weitzenb{\"{o}}ck
connection:
\begin{align}
&  \!\!\!\!\!\!\!\!\!\! T^{(W)}=\frac{1}{4} \left(  W^{\mu\lambda\nu}-
W^{\mu\nu\lambda} \right)  \left(  W_{\mu\lambda\nu} -W_{\mu\nu\lambda}\right)
\nonumber\\
&  \ \ \ \ \ +\frac{1}{2} \left(  W^{ \mu\lambda\nu} -W^{ \mu\nu\lambda}
\right)  \left(  W_{\lambda\mu\nu} -W_{\lambda\nu\mu}\right) \nonumber\\
&  \ \ \ \ \ - \left(  W_{\nu}^{\,\,\,\mu\nu} -W_{\nu}^{\,\,\,\nu\mu}\right)
\left(  W^{\lambda}_{\,\,\,\mu\lambda}-W^{\lambda}_{\,\,\,\lambda\mu}\right)
. \label{TdefW}%
\end{align}

In general relativity one uses $R^{(LC)}$ in the Lagrangian,
while in teleparallel
equivalent of general relativity ones uses $T^{(W)}$.
Both these theories possess two propagating degrees of 
freedom, describing   a massless spin-two field, i.e. the graviton. 
Thus, in their corresponding modifications, namely 
curvature-based modified gravity or torsion-based modified theories, one can 
acquire extra  degrees of freedom by extending the action, and these  extra  
degrees of freedom are the ones that lead to modified cosmological evolution.
Nevertheless, after the above discussion we realize that one can 
 introduce extra degrees of freedom through the consideration of non-special 
connections, i.e.  going beyond the  
Levi-Civita and  Weitzenb{\"{o}}ck ones. Hence, if ones applies a 
connection that has both non-zero curvature and torsion,   he   
obtains a theory with more degrees of freedom.

Specifically, as it was presented in \cite{Myrzakulov:2012qp,Saridakis:2019qwt} 
one can
construct a theory that is based on a specific but not special connection that
leads to both non-zero curvature and non-zero torsion.   The action of
such a theory would be
\begin{equation}
S = \int d^{4}x e \left[  \frac{F(R,T)}{2\kappa^{2}} +L_{m} \right]  ,
\label{action1}%
\end{equation}
with $e = \text{det}(e_{\mu}^{a}) = \sqrt{-g}$ and $\kappa^{2}=8\pi G$ the
gravitational constant, however we mention that $T$ and $R$ are the torsion
and curvature scalars of the non-special connection, namely
\cite{Kofinas:2014owa}
\begin{align}
&  T=\frac{1}{4}T^{\mu\nu\lambda}T_{\mu\nu\lambda}+\frac{1}{2}T^{\mu\nu
\lambda} T_{\lambda\nu\mu}-T_{\nu}^{\,\,\,\nu\mu}T^{\lambda}_{\,\,\,\lambda
\mu},\label{Tdef2}\\
&  R=R^{(LC)}+T-2T_{\nu\,\,\,\,\,\,\,\,;\mu}^{\,\,\,\nu\mu}\,,
\label{Radef222}%
\end{align}
with $;$ denoting the covariant differentiation with respect to the
Levi-Civita connection. Finally, in the above action we have also added the
matter Lagrangian $L_{m}$.

As one can see from the definitions (\ref{curvaturebastard}%
),(\ref{torsionbastard}) $T$ depends on the tetrad, its first derivative and
the connection, and $R$ depends on the tetrad and its first and second
derivatives, and on the connection and its first derivative. These allows one
to introduce the parametrization \cite{Saridakis:2019qwt}
\begin{align}
&  T=T^{(W)}+v,\label{T1}\\
&  R=R^{(LC)} + u, \label{R1}%
\end{align}
with $u$ being a scalar quantity depending on the tetrad, its first and second
derivatives, and the connection and its first derivative, and $v$ a scalar
depending on the tetrad, its first derivative and the connection.

The above theory has non-trivial structure and exhibits extra degrees of
freedom even in the case where the arbitrary function $F(R,T)$ has a trivial
form, since the novel features arise from the non-trivial connection itself,
parametrized by the quantities $u$ and $v$. If this connection becomes the
Levi-Civita one, we obtain that $u=0$ and $v=-T^{(W)}$, and thus we recover
the standard $F(R)$ gravity (which for $F(R)=R$ becomes general relativity).
However, if the connection is the Weitzenb{\"{o}}ck one, then we acquire $v=0$
and $u=-R^{(LC)}$ and therefore we recover standard $F(T)$ gravity (which for
$F(T)=T$ becomes the teleparallel equivalent of general relativity).

In order to proceed to the cosmological applications of the above
construction, we follow the mini-super-space procedure
\cite{Saridakis:2019qwt}. Imposing the flat Friedmann-Robertson-Walker (FRW)
metric
\begin{align}
ds^{2}= dt^{2}-a^{2}(t)\, \delta_{ij} dx^{i} dx^{j},
\end{align}
namely the tetrad $e^{a}_{\,\,\,\mu}=\mathrm{diag}[1,a(t),a(t),a(t)]$, with
$a(t)$ the scale factor, we find $R^{(LC)}=6 \left(  \frac{\ddot{a}}{a}+
\frac{\dot{a}^{2}}{a^{2}}\right)  $ and $T^{(W)}=-6 \left(  \frac{\dot{a}^{2}%
}{a^{2}} \right)  $. Taking into account the dependence of $u$ and $v$ on the
metric and the connection, we deduce that $u=u(a,\dot{a},\ddot{a})$ and
$v=v(a,\dot{a})$. Furthermore, we take the standard form $L_{m}=-\rho_{m}(a)$
\cite{Dimakis:2016mip}. Lastly, in order to explore the
dynamics of Myrzakulov gravity arising solely form the non-special connection
itself, we make the simple linear choice $F(R,T)=R+\lambda T$, with $\lambda$
the dimensionless coupling parameter.

Inserting the above mini-super-space expressions into (\ref{action1}) we
obtain $S=\int Ldt$, with
\begin{align}
&  \!\!\!\!\!\!\!\!\!\!\! L= \frac{3}{\kappa^{2}}\left[  \lambda+1\right]
a\dot{a}^{2}- \frac{ a^{3}}{2\kappa^{2}}\left[  u(a,\dot{a},\ddot{a})+\lambda
v(a,\dot{a}) \right] \nonumber\\
&  \, + a^{3} \rho_{m}(a) . \label{4.2}%
\end{align}
We can now perform variation and extract the equations of motion for $a$, and
we can moreover consider the Hamiltonian constraint $\mathcal{H}=\dot
{a}\left[  \frac{\partial L}{\partial\dot{a}}- \frac{\partial}{\partial
t}\frac{\partial L}{\partial\ddot{a}}\right]  +\ddot{a}\left(  \frac{\partial
L}{\partial\ddot{a}}\right)  -L=0 $. Hence, we result to the following
Friedmann equations \cite{Saridakis:2019qwt}
\begin{align}
3H^{2}  &  = \kappa^{2}\left(  \rho_{m}+\rho_{MG} \right) \label{FR1a}\\
2\dot{H}+3H^{2}  &  = -\kappa^{2} \left(  p_{m}+ p_{MG}\right)  , \label{FR2a}%
\end{align}
where the dark energy sector that arises effectively from the non-special
connection has energy density and pressure
\begin{align}
&  \!\!\!\!\!\!\!\!\!\!\!\!\!\!\!\! \rho_{MG}=\frac{1}{\kappa^{2}}
\Big[ \frac{Ha}{2} \left(  u_{\dot{a}}+v_{\dot{a}} \lambda\right)  -\frac
{1}{2} (u+\lambda v)\nonumber\\
&  \ \ \ \ \ \ \ + \frac{a u_{\ddot{a}}}{2} \left(  \dot{H}-2 H^{2}\right)
-3\lambda H^{2}\Big]\label{rhoDEa1}\\
&  \!\!\!\!\!\!\!\!\!\!\!\!\!\!\!\! p_{MG}= -\frac{1}{\kappa^{2}}
\Big[\frac{Ha}{2} \left(  u_{\dot{a}}+v_{\dot{a}} \lambda\right)  -\frac{1}{2}
(u+ \lambda v)\nonumber\\
&  \ \ \ \ \ \ \ \ \ -\frac{a}{6} \left(  u_{a}+\lambda v_{a}-{\dot{u}%
_{\dot{a}}}-\lambda{\dot{v}_{\dot{a}}}\right) \nonumber\\
&  \ \ \ \ \ \ \ \ \ -\frac{a}{2}\left(  \dot{H}+3H^{2}\right)  u_{\ddot{a}}-H
a \dot{u}_{\ddot{a}}\nonumber\\
&  \ \ \ \ \ \ \ \ \ -\frac{a}{6} \ddot{u}_{\ddot{a}} -\lambda(2\dot{H}%
+3H^{2})\Big], \label{pDEa1}%
\end{align}
respectively. In the above expressions $H=\frac{\dot{a}}{a}$ is the Hubble
parameter, $p_{m}$ is the pressure of the matter sector, and the subscripts
$a,\dot{a},\ddot{a}$ mark partial derivatives with respect to these arguments.
Note that the effective dark energy sector is conserved, namely $\dot{\rho
}_{MG}+3H(\rho_{MG}+p_{MG})=0$, as it is easily deduced from the above
imposing the matter conservation equation $\dot{\rho}_{m}+3H(\rho_{m}%
+p_{m})=0$ too.

In the following we focus on two classes of the theory at hand, constructed 
phenomenologically in order to lead to interesting cosmological evolution.

\subsection{Class 1}

\label{class1def}

As a first example we consider the class where $u=c_{1}\dot{a}-c_{2}$ and
$v=c_3\dot{a}-c_{4}$, where $c_{1}$,$c_{2}$,$c_{3}$,$c_{4}$ are constants. For
this class equations (\ref{FR1a})-(\ref{pDEa1}) lead to
\begin{align}
3H^{2}  &  =\kappa^{2}\left(  \rho_{m}+\rho_{MG}\right) \label{FR1a1}\\
2\dot{H}+3H^{2}  &  =-\kappa^{2}\left(  p_{m}+p_{MG}\right)  , \label{FR2a1b}%
\end{align}
with
\begin{align}
&  \rho_{MG}=\frac{1}{\kappa^{2}}\left[  c-3\lambda H^{2}\right]
\label{rhoDEa}\\
&  p_{MG}=-\frac{1}{\kappa^{2}}\left[  c-\lambda(2\dot{H}+3H^{2})\right]  ,
\label{pDEa}%
\end{align}
where we have defined $c\equiv c_{2}+c_{4}$. Thus, the effective dark-energy
equation-of-state parameter reads
\begin{equation}
w_{DE}=-1+\frac{2\lambda\dot{H}}{c-3\lambda H^{2}}. \label{wmgclass1}%
\end{equation}
It is interesting to  mention here that the effective dark energy density 
(\ref{rhoDEa}) falls within particular  subclasses of the running vacuum 
cosmology \cite{rvm1,rvmevol,solaqft}.

\subsection{Class 2}

\label{class2def}

The second class that we are interested in is the one characterized by
$u=c_{1}\frac{\dot{a}}{a}\ln\dot{a}$ and $v=s(a)\dot{a}$, where $s(a)$ is an
arbitrary function. Hence, expressions (\ref{FR1a})-(\ref{pDEa1}) give
again the Friedmann equations
(\ref{FR1a1}),(\ref{FR2a1b})
but now   
with
\begin{align}
&  \rho_{MG}=\frac{1}{\kappa^{2}}\left[  \frac{c_{1}}{2}H-3\lambda
H^{2}\right]  \label{rhoDEb}\\
&  p_{MG}=-\frac{1}{\kappa^{2}}\left[  \frac{c_{1}}{2}H+\frac{c_{1}}{6}%
\frac{\dot{H}}{H}-\lambda(2\dot{H}+3H^{2})\right]  ,\label{pDEb}%
\end{align}
and thus we can find
\begin{equation}
w_{DE}=-1+\frac{2\lambda\dot{H}-\frac{c_{1}}{6}\frac{\dot{H}}{H}}{\frac{c_{1}%
}{2}H-3\lambda H^{2}}.\label{wmgclass2}%
\end{equation}
Similarly to the Class 1 above,   the effective dark energy density 
(\ref{rhoDEb}) coincides 
with broader subclasses of the running vacuum cosmology,
and as we show below it can lead to very interesting cosmological behavior 
despite the fact that it does not have $\Lambda$CDM scenario as a 
particular limit.

\section{Phase Space Analysis}

\label{analysis}

In the previous section we presented the cosmological equations of the
scenario at hand. As we can see, Myrzakulov gravity leads to the appearance of
new terms in the Friedmann equations, that are of geometrical origin and in
particular they arise from the non-trivial connection structure through the
parametrization in terms of $u$ and $v$. In this section we proceed to the
full phase-space analysis of these scenarios, by applying the dynamical system
method \cite{REZA,Coley:2003mj}. Hence, we will first introduce suitably the
auxiliary variables needed in order to transform the equations into an
autonomous dynamical system
\cite{REZA,Coley:2003mj,Copeland:1997et,Leon:2010pu,
Fadragas:2013ina,Paliathanasis:2015gga,Carloni:2015bua,
Nersisyan:2016hjh,Papagiannopoulos:2017whb, 
SantosDaCosta:2018bbw,Basilakos:2019dof, Bahamonde:2019urw, Khyllep:2021wjd}, 
and then we will extract
its critical points. Thus, examining the eigenvalues of the perturbation
matrix around each of them we can conclude on their stability properties.

In order to perform the dynamical analysis we introduce the quantities
\begin{align}
&  A= \Big[ \frac{Ha}{2} \left(  u_{\dot{a}}+v_{\dot{a}} \lambda\right)
-\frac{1}{2} (u+\lambda v)\nonumber\\
&  \ \ \ \ \ \ \ \ + \frac{a u_{\ddot{a}}}{2} \left(  \dot{H}-2 H^{2}\right)
\Big],
\nonumber\\
&  B=\Big[ -a \left(  u_{a}+\lambda v_{a}-{\dot{u}_{\dot{a}}}-\lambda{\dot
{v}_{\dot{a}}}\right) \nonumber\\
&  \ \ \ \ \ \ \ \ \ -3a\left(  \dot{H}+3H^{2}\right)  u_{\ddot{a}}-6H a
\dot{u}_{\ddot{a}}\nonumber\\
&  \ \ \ \ \ \ \ \ \ -a \ddot{u}_{\ddot{a}} \Big]. \label{ABdefinit}%
\end{align}
Hence, the two Friedman equations can be written as
\begin{equation}
3H^{2}(1+\lambda)=\kappa^{2}\rho_{m}+A\ \ \label{fr11}%
\end{equation}
\begin{equation}
(2\dot{H}+3H^{2})(1+\lambda)=\kappa^{2}\rho_{m}w_{m}+A+\frac{B}{6},
\label{fr22}%
\end{equation}
where for convenience we have also introduced the matter equation-of-state
parameter defined as $w_{m}=p_{m}/\rho_{m}$.

Let us first examine the limit of the scenario at hand to the $\Lambda$CDM
cosmology. In order to achieve this we need $\rho_{MG}=-p_{MG}$, which implies
that
\begin{equation}
\lambda\dot{H}=-\frac{B}{12}. \label{con11}%
\end{equation}
Although this condition can be satisfied in many ways, the simplest one is to
consider the case $\lambda=0$, namely to focus on a Lagrangian being just the
curvature $R$ corresponding to the non-special connection. In this case, if we
choose a connection with $u=c_{1} \dot{a}-c_{2}$, where $c_{1}$,$c_{2}$ are
constants, we acquire
\begin{equation}
\rho_{MG}=-p_{MG}=\frac{c_{2}}{2\kappa^{2}}\equiv\Lambda.
\end{equation}
Interestingly enough, we observe that we do obtain $\Lambda$CDM cosmology
although in the starting action we had not considered an explicit cosmological
constant. Thus, the non-trivial structure of the underlying geometry results
to an effective cosmological constant, which reveals the capabilities of the
theory. Note that even in this simple case where $\lambda=0$, and thus $T$
disappears from the action, the non-special connection still has a non-zero
torsion. In general, such an effective emergence of a cosmological constant
due to the richer underlying connection appears in other geometrical modified
gravities too \cite{Ikeda:2019ckp,Minas:2019urp}, and reveals the advantages
of the theory.

Having the above discussion in mind we can deduce that Class 1 defined in
subsection \ref{class1def} corresponds to a deviation from $\Lambda$CDM
cosmology, accepting it as a particular limit and thus satisfying the basic
requirements to be a viable theory, while still maintaining the possibility to
improve $\Lambda$CDM behavior. On the other hand, Class 2 defined in
subsection \ref{class2def} does not have $\Lambda$CDM cosmology as a limit,
nevertheless, and interestingly enough, as we will later show it can lead to a
cosmological behavior in agreement with observations.

We can now proceed to the dynamical analysis of the above specific classes,
keeping a general $\lambda\neq0$.

\subsection{Class 1}

We start with Class 1 of subsection \ref{class1def}. In this case definitions
(\ref{ABdefinit}) lead to
\begin{align}
A  &  =\frac{1}{2}(c_{2}+\lambda c_{4})\equiv C\nonumber \\
B  &  =0.
\end{align}
In order to transform the cosmological equations into an autonomous form we
introduce the dark matter and dark energy density parameters as our
dimensionless variables, namely
\begin{align}
&  \Omega_{m}\equiv \frac{\kappa^{2}\rho_{m}}{3H^{2}(1+\lambda)}\label{Omm1}\\
&  \Omega_{MG}\equiv \frac{C}{3H^{2}(1+\lambda)}, \label{Ode1}%
\end{align}
and therefore the first Friedmann equation (\ref{FR1a1}) becomes $1=\Omega
_{m}+\Omega_{MG}$ (note that the case $\lambda=-1$ is not physically
interesting since according to (\ref{fr11}) leads to $\rho_{m}=-C/\kappa
^{2}=const.$, and hence in the following we focus on the case $\lambda\neq
-1$).
Additionally, the second  Friedmann equation (\ref{FR2a1b})  
becomes
\begin{equation}
\frac{\dot{H}}{H^{2}}=-\frac{3}{2}(1+\Omega_{m}w_{m}-\Omega_{MG}).
\end{equation}
Using this expression, as well as (\ref{Omm1}),(\ref{Ode1}), equation 
(\ref{wmgclass1})
can be rewritten as follows:%
\begin{equation}
 w_{DE}\mathbf{=\ }-1-\frac{3\lambda(1+\Omega_{m}w_{m}-\Omega_{MG}%
)}{\Omega_{MG}(1+\lambda)-3\lambda}\mathbf{.}\label{wmg1_dim}%
\end{equation}

In summary, the   dynamical system   can be straightforwardly written
as:
\begin{eqnarray}
&&
\frac{d\Omega_{m}}{d\ln a}=-3\Omega_{m}\left[  w_{m}-\ \Omega_{m}w_{m}%
+\Omega_{MG}\right],
\\
&&
\frac{d\Omega_{MG}}{d\ln a}=3\Omega_{MG}\left[  \ \Omega_{m}w_{m}%
+1-\Omega_{MG}\right]  .
\end{eqnarray} 

Since the first Friedman equation acts as a constraint, we finally remain with 
one-dimensional phase space. The corresponding critical points $P(\Omega_{m}
,\Omega_{MG})$ are summarized in Table \ref{taba011}, alongside their features 
and    stability conditions. Note that in this case (\ref{wmg1_dim}) provides 
the useful expression
\begin{equation}
 w_{DE}\mathbf{=\ }-1-\frac{3\lambda \Omega_{m}   
(1+w_{m})}
{(1-\Omega_{m})(1+\lambda)-3\lambda}.\label{wmg1_dim2}
\end{equation}

\begin{table}[ht] \centering
\resizebox{\columnwidth}{!}
{
\begin{tabular}
[c]{cccccc}\hline\hline
\textbf{Point} & $(\Omega_{m},\Omega_{MG})$ & \textbf{Existence} &
$w_{DE}$ & \textbf{Acceleration} & \textbf{Stability}\\\hline
$P_{1}$ & $\left(  0,1\right)  $ & Always & $-1$ & yes & $w_{m}>-1$\\
$P_{2}$ & $(1,0)$ & Always & $w_{m}$ & $w_{m}<-\frac{1}{3}$ & $w_{m}%
<-1$\\\hline\hline
\end{tabular}}
\caption{The physically interesting critical points of Class 1, namely of 
(\ref{FR1a1}),(\ref{FR2a1b}) with (\ref{rhoDEa}),(\ref{pDEa}),
  their features and their
  stability conditions.}%
  \label{taba011} 
\end{table}

As we observe, point $P_1$ corresponds to a dark-energy dominated
Universe, in which dark energy behaves like a cosmological constant, and the 
fact that in the usual case of dust matter  it is stable implies that it will 
be the late-time state of the Universe independently of the initial conditions. 
On the other hand, point $P_2$ is a matter-dominated, non-accelerating 
solution, and the fact that for dust matter it is saddle implies that this 
point can describe the necessary intermediate era of the Universe, in which 
matter structure is formed \cite{Basilakos:2019dof,Khyllep:2021wjd}.

\begin{figure}[ht]
	\centering
		\includegraphics[width=0.45\textwidth]{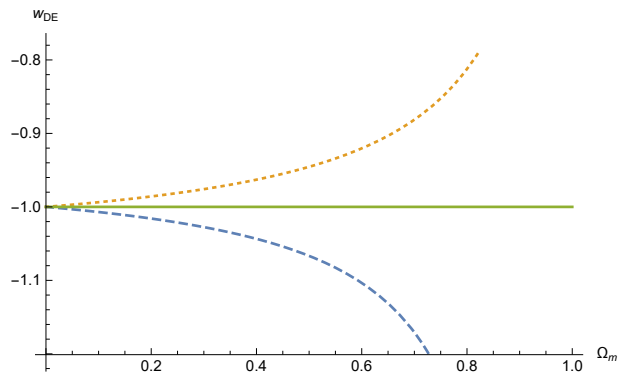}
	\caption{{\it{The behavior of the system in the $(w_{DE},\Omega_m)$ space
	 for  Class 1 models   of 
(\ref{FR1a1}),(\ref{FR2a1b}) with (\ref{rhoDEa}),(\ref{pDEa}), for $w_m=0$ and 
with  $\lambda=0.02$ (blue-dashed),   $\lambda=0$ (green-solid) and  
$\lambda=-0.02$ (orange-dotted).}}}
	\label{case1wOm}
\end{figure}

\begin{figure}
	\centering
		\includegraphics[width=0.45\textwidth]{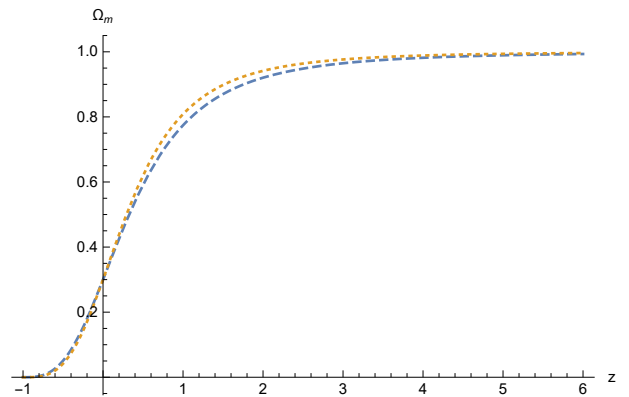}
	\caption{{\it{The evolution of the matter density parameter $\Omega_{m}(z)$ 
as a function of the redshift, for  Class 1 models   of 
(\ref{FR1a1}),(\ref{FR2a1b}) with (\ref{rhoDEa}),(\ref{pDEa}), for $w_m=0$  
(blue-dashed) and 
   $w_m=0.1$    (orange-dotted). }}
	 }
	\label{fig:Om_plots_casea}
\end{figure}

In order to show the above feature in a more transparent way, in Fig. 
\ref{case1wOm} we 
present the behavior of the system in the $(w_{DE},\Omega_m)$ space, in the 
case of dust matter, for various values of $\lambda$. As we can see, the 
system passes through the saddle point 
$P_2$ before it results to the stable late-time attractor $P_1$. Additionally, 
 in order to examine the system at both intermediate and 
late times,  in Fig. \ref{fig:Om_plots_casea} we present $\Omega_{m}$ as a 
function of the redshift  
$z=-1+a_0/a$ (setting the current scale factor $a_0=1$),   since  $\dot{H}= 
-(1+z)H(z)H'(z)$ with primes denoting 
derivatives with respect to $z$. We choose different values of $w_m$, and 
we fix  $C$ in order 
to have $\Omega_{m}(z=0)\equiv\Omega_{m0}\approx0.31$ as required by 
observations \cite{Planck:2018vyg}. 
As we observe, the Universe follows the required evolution, with the sequence 
of matter and dark energy epochs.

\begin{figure}
	\centering
		\includegraphics[width=0.45\textwidth]{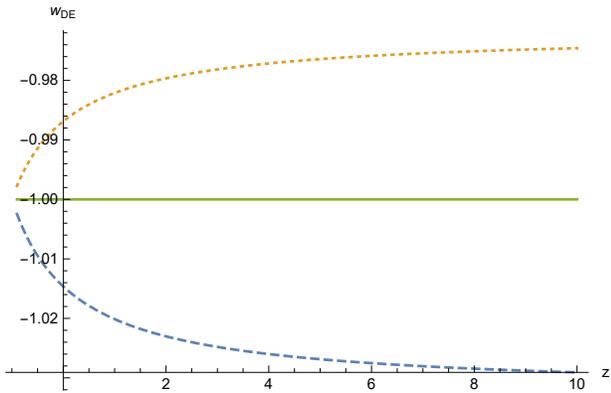}
	\caption{{\it{The evolution of the  dark-energy equation-of-state 
parameter $w_{DE}(z)$  
as a function of the redshift, for  Class 1 models   of 
(\ref{FR1a1}),(\ref{FR2a1b}) with (\ref{rhoDEa}),(\ref{pDEa}), for $w_m=0$ and 
with  $\lambda=0.02$ (blue-dashed), $\lambda=0$ (green-solid) and  
$\lambda=-0.02$ (orange-dotted).}}}
	\label{fig:Wde_plots casea1}
\end{figure}

Moreover, in Fig. \ref{fig:Wde_plots casea1} 
we depict the corresponding behavior of the dark-energy equation-of-state 
parameter $w_{DE}$ for various values of $\lambda$. As we see, although for 
every $\lambda$ at asymptotic late times (i.e. for $z\rightarrow-1$) $w_{DE}$ 
is stabilized at the cosmological constant value $-1$, as it was found in Table  
\ref{taba011}, the behavior at intermediate redshifts and at the present 
Universe is different. In particular, for $\lambda<0$ the dark energy sector 
behaves as quintessence, while for $\lambda>0$ the $w_{DE}$  lies in the 
phantom regime. This was expected from the form of (\ref{wmg1_dim2}), and 
reveals that Class 1 offers a unified description of both quintessence 
and phantom regimes, without pathologies. Finally, as we see, in the case 
$\lambda=0$ the scenario at hand recovers $\Lambda$CDM cosmology.

\subsection{Class 2}

Let us now proceed to the investigation of Class 2 of subsection
\ref{class2def}. In this case definitions (\ref{ABdefinit}) lead to
\begin{eqnarray}
 A=\frac{c_{1}}{2}H\ \equiv DH.
\end{eqnarray} 
Similarly to the Class 1 case, for $\lambda\neq-1$ we can introduce the
dimensionless auxiliary variables
\begin{align}
&  \Omega_{m}\equiv \frac{\kappa^{2}\rho_{m}}{3H^{2}(1+\lambda)}\label{Omm2}\\
&  \Omega_{MG}\equiv \frac{D}{3H(1+\lambda)}, \label{Ode2}
\end{align}
and thus the first Friedmann equation   becomes the constraint 
$1=\Omega
_{m}+\Omega_{MG}$. Additionally,  for $\lambda\neq-1$ the second Friedmann 
equation   becomes
\begin{eqnarray}
\!\!\!\!\!\!\!\!\!\!\!\!\!\frac{\dot{H}}{H^{2}}=-\frac{3(1+\Omega_{m}%
w_{m}-\Omega_{MG})}{2- \Omega_{MG}}.
\end{eqnarray} 
Hence, using this expression and (\ref{Omm2}),(\ref{Ode2}) we can rewrite 
expression (\ref{wmgclass2}) as
\begin{equation}
w_{DE}=-1-\frac{(1+\Omega_{m}w_{m}-\Omega_{MG})  [\lambda(2-\Omega_{MG} 
)-\Omega_{MG}
]}{(2-\Omega_{MG})[(1+\lambda)\Omega
_{MG}-\lambda]}.
\label{wmg2_dim}
\end{equation}

For this class of scenarios, the dynamical system can be
straightforwardly written as:
\begin{eqnarray}
 &&\!\!\!\!\!\!\! \!\!\!\frac{d\Omega_{m}}{d\ln a}=\frac
{3\Omega_{m} \left[  \Omega_{MG}-w_{m}(-2+\Omega_{MG}+2\Omega_{m})\right]
}{\Omega_{MG}-2},\\
&&\!\!\!\!\!\!\!\!\!\!
 \frac{d\Omega_{MG}}{d\ln a}=\frac{6\Omega_{MG} (\Omega_{MG}-w_{m}\Omega
_{m}-1)}{\Omega_{MG}-2}.
\end{eqnarray}
 Due to the constraint first Friedman equation, we 
result to a one-dimensional phase space. Hence, 
 in this case (\ref{wmg2_dim}) gives 
the useful expression
\begin{equation}
w_{DE}=-1-\frac{ \Omega_{m} (1+w_{m} )  
[\lambda-1 + \Omega_{m}(\lambda+1 )]}
{(1+\Omega_{m})[(1+\lambda) (1-\Omega_{m}) -\lambda]}.
\label{wmg2_dimb}
\end{equation}

\begin{table}[ht] \centering
\resizebox{\columnwidth}{!}
{
\begin{tabular}
[c]{cccccc}\hline\hline
\textbf{Point} & $(\Omega_{m},\Omega_{MG})$ & \textbf{Existence} &
$w_{DE}$ & \textbf{Acceleration} & \textbf{Stability}\\\hline
$P_{1}$ & $\left(  0,1\right)  $ & Always & $-1$ & yes & $w_{m}>-1$\\
$P_{2}$ & $(1,0)$ & Always & $w_{m}$ & $w_{m}<-\frac{1}{3}$ & $w_{m}%
<-1$\\\hline\hline
\end{tabular}}
\caption{The physically interesting critical points of Class 2, namely of 
(\ref{FR1a1}),(\ref{FR2a1b}) with (\ref{rhoDEb}),(\ref{pDEb}),
  their features and their
  stability conditions.  }%
\label{taba01bb}
\end{table}
 \begin{figure}[ht]
	\centering
		\includegraphics[width=0.45\textwidth]{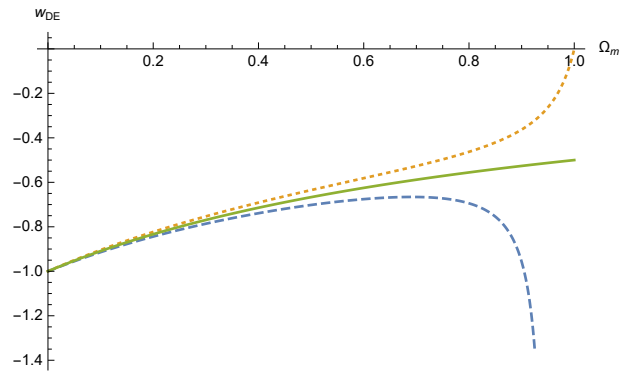}
	\caption{{\it{The behavior of the system in the $(w_{DE},\Omega_m)$ space
	 for  Class 2 models   of 
(\ref{FR1a1}),(\ref{FR2a1b}) with (\ref{rhoDEb}),(\ref{pDEb}), for $w_m=0$ and 
with  $\lambda=0.02$ (blue-dashed),   $\lambda=0$ (green-solid) and  
$\lambda=-0.02$ (orange-dotted).}}}
	\label{case2wOm}
\end{figure}

The   critical 
points  are summarized in Table \ref{taba01bb}. In the same Table we provide 
 their features and  their stability conditions.
 Interestingly enough, Class 2 exhibits the same critical points with Class 1, 
namely the dark-energy dominated, de Sitter Universe  $P_1$, which is stable 
for dust matter, and the matter-dominated, non-accelerating Universe $P_2$, 
which is saddle for dust matter. The importance of the current behavior is that 
it is obtained not only without the consideration of an explicit cosmological 
constant, but also through the quite rich and complicated  dark energy 
density (\ref{rhoDEb}), which does not accept $\Lambda$CDM model as a 
particular limit.
\begin{figure}[ht]
	\centering
		\includegraphics[width=0.45\textwidth]{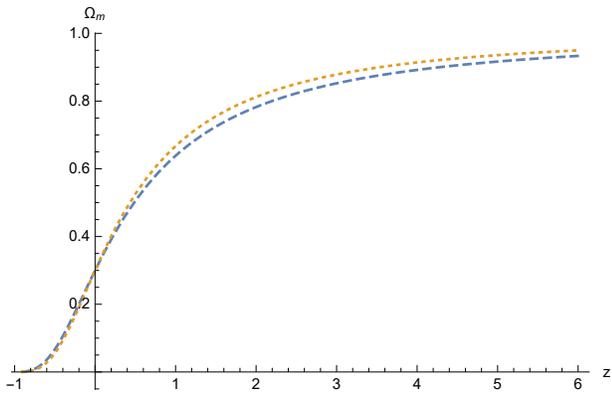}
	\caption{{\it{The evolution of the matter density parameter $\Omega_{m}(z)$ 
as a function of the redshift, for  Class 2 models   of 
(\ref{FR1a1}),(\ref{FR2a1b}) with (\ref{rhoDEb}),(\ref{pDEb}), for $w_m=0$  
(blue-dashed) and 
   $w_m=0.1$    (orange-dotted). }}}
	\label{fig:Om_plots_caseb}
\end{figure}
\begin{figure}[ht]
	\centering
		\includegraphics[width=0.45\textwidth]{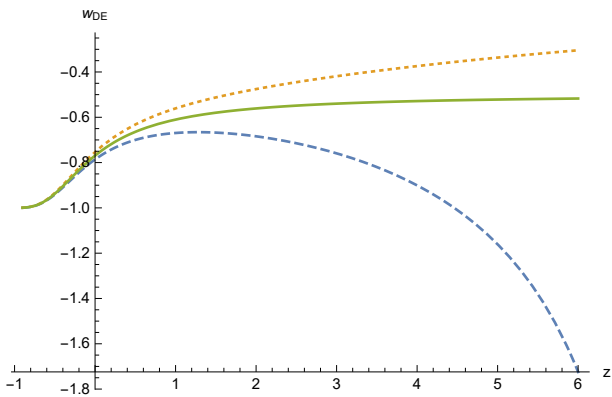}
	\caption{{\it{The evolution of the  dark-energy equation-of-state 
parameter $w_{DE}(z)$  
as a function of the redshift, for  Class 2 models   of 
(\ref{FR1a1}),(\ref{FR2a1b}) with (\ref{rhoDEb}),(\ref{pDEb}), for $w_m=0$ and 
with  $\lambda=0.02$ (blue-dashed), $\lambda=0$ (green-solid) and  
$\lambda=-0.02$ (orange-dotted).}} }
	\label{fig:Wde_plots caseb1}
\end{figure}

Nevertheless, although Class 2 has the same critical points with Class 1, the 
behavior of the system at intermediate times is radically different. In Fig. 
\ref{case2wOm}
we show the  $(w_{DE},\Omega_m)$ diagram, in the 
case of dust matter. Although the system passes through the saddle point 
$P_2$ before it results to the stable late-time attractor $P_1$, the 
corresponding curves are different from those of Fig \ref{case1wOm}. 
Furthermore, in Fig.  \ref{fig:Om_plots_caseb} we present $\Omega_{m}$ as a 
function of the redshift $z$   and fixing $D$ in 
order 
to have $\Omega_{m}(z=0)\equiv\Omega_{m0}\approx0.31$, 
where we can see 
the  sequence of matter and dark energy eras.

Finally, in Fig. \ref{fig:Wde_plots caseb1} 
we present the evolution of  the   dark-energy 
equation-of-state 
parameter $w_{DE}(z)$ for various values of $\lambda$. 
Similarly to the previous class of models, at asymptotically late times  
$w_{DE}$  tends to the cosmological constant value $-1$, as it was found in 
Table \ref{taba01bb}, however in the present case this is not trivial since the 
scenario at hand does not possess $\Lambda$CDM cosmology as a limit.
Additionally,  the behavior at intermediate redshifts is even more different, 
and the dark-energy sector can lie in the quintessence regime, in the phantom 
regime, or experience the phantom-divide crossing during the evolution.
 This was expected from the form of (\ref{wmg2_dimb}), and 
reveals the capabilities of this Class of models.

\section{Discussion}
\label{Conclusions}

We performed a dynamical   system analysis of Myrzakulov  
 gravity. The latter is a subclass of affinely connected metric theories, in 
which  ones uses a specific but non-special connection, which allows for 
non-zero curvature and torsion simultaneously. Thus, one obtains
extra degrees of freedom which in turn lead to extra terms in the Friedman 
equations that can lead to interesting phenomenology. Hence, by applying the 
dynamical system approach and performing a phase-space analysis, one is able to 
 bypass the non-linearities of the equations and investigate the  global 
behavior of the system,  independently of the specific
initial conditions or the intermediate-time behavior evolution.

We considered two classes of models and for each case we transformed the 
equations into an autonomous dynamical system. We extracted the critical 
points, and we examined their stability properties alongside their physical 
features. In the Class 1 models, which possess $\Lambda$CDM cosmology as a 
limit, we found  that independently of the 
initial conditions the Universe will result in a dark-energy dominated 
critical point in  which dark energy behaves like a cosmological constant.
Moreover, we found a matter-dominated, non-accelerating 
solution, which is saddle and thus it can describe the necessary corresponding 
intermediate matter 
era of the Universe. Hence,   the Universe follows the required evolution, with 
the sequence 
of matter and dark energy eras. Concerning the   dark-energy 
equation-of-state parameter $w_{DE}$,   we   showed that   although   
 at asymptotic late times  it is stabilized at the cosmological constant value 
$-1$ for every value of the model parameter $\lambda$,
 the behavior at intermediate redshifts and at the present 
Universe is different, since for $\lambda<0$ the dark energy 
sector behaves as quintessence, while for $\lambda>0$ the $w_{DE}$  lies in the 
phantom regime.  

For the Class 2 models, we again found the dark-energy dominated, de Sitter 
late-time attractor, and the saddle critical point corresponding to  
matter-dominated, non-accelerating Universe. Furthermore,  at asymptotically 
late times  
$w_{DE}$  tends to the cosmological constant value $-1$.
However, the interesting feature is that 
this was obtained without the scenario   possessing $\Lambda$CDM cosmology as a 
particular limit. This Class can also describe the sequence of of matter and 
dark energy epochs, nevertheless the  at intermediate times the behavior is 
radically different than the previous Class, since   the dark-energy 
sector can lie in the quintessence regime, in the phantom 
regime, or experience the phantom-divide crossing during the evolution.

Let us stress here that, as we mentioned above, the two examined classes 
of theories, at a cosmological framework, fall withing the class of generalized 
running vacuum theories  \cite{rvm1,rvmevol,solaqft}. Hence, one can perform 
the Big-Bang Nucleosynthesis analysis in the same way \cite{Asimakis:2021yct} 
and deduce that the early-universe evolution is not spoiled in the present 
models, too.
 
In summary, the phase-space analysis revealed the interesting features of 
Myrzakulov    gravity, and in particular the ability to posses a 
stable de Sitter solution as a late-time attractor even without the explicit 
consideration of a cosmological constant. It would be interesting to apply the 
Noether symmetry approach \cite{Paliathanasis:2014iva} in order to extract exact
analytic solutions at intermediate times too. Furthermore, since the resulting 
cosmological equations are similar to subclasses of the running vacuum 
cosmology, it is necessary to further investigate their possible connection, 
and examine whether the current framework offers the way to provide a 
Lagrangian for running vacuum models, a well-known open issue in the 
corresponding literature. Finally, it would be interesting to investigate 
the relation and differences of the present theory with theories with Weyl 
connection (not to be confused with Weyl gravity, that uses the standard 
Levi-Civita connection), which have   an altered connection but non-zero 
non-metricity  \cite{Gomes:2018sbf}.These
studies will be performed in  
separate works.

\begin{acknowledgments}
GP is supported by the scholarship of the Hellenic Foundation for Research and
Innovation (ELIDEK grant No. 633). SB acknowledges support by the Research
Center for Astronomy of the Academy of Athens in the context of the program
``Testing general relativity on cosmological scales'' (ref. number 200/872). 
ENS  acknowledges participation in the COST Association Action CA18108 
``{\it Quantum Gravity Phenomenology in the Multimessenger Approach (QG-MM)}''.
\end{acknowledgments}

\end{document}